\documentclass[conference]{IEEEtran}

\usepackage{geometry}
\usepackage[normalem]{ulem}
\usepackage{amsmath}
\usepackage{comment}
\usepackage{amsfonts}
\usepackage{enumitem}
\usepackage{tablefootnote}
\usepackage{cite}
\usepackage{xcolor}
\usepackage{algorithm} 
\usepackage{algpseudocode} 
\usepackage{bm}
\usepackage{threeparttable, tablefootnote}
\usepackage{bbm}
\usepackage[nice]{nicefrac}
 
\usepackage[acronyms,nonumberlist,nopostdot,nomain,nogroupskip]{glossaries}

\makeatletter
\newcommand{\algmargin}{\the\ALG@thistlm}   
\makeatother
\algnewcommand{\parState}[1]{\State%
    \parbox[t]{\dimexpr\linewidth-\algmargin}{\strut #1\strut}}

\ifCLASSINFOpdf
\usepackage[pdftex]{graphicx}
\else
\fi

\ifCLASSOPTIONcompsoc
  \usepackage[caption=false,font=normalsize,labelfont=sf,textfont=sf]{subfig}
\else
  \usepackage[caption=false,font=footnotesize]{subfig}
\fi

\newcounter{myfootertablecounter}

\usepackage{soul}

\newacronym{3gpp}{3GPP}{3rd Generation Partnership Project}
\newacronym{5g}{5G}{5th generation}
\newacronym{5gc}{5GC}{5G Core}
\newacronym{6g}{6G}{6th generation}
\newacronym{ack}{ACK}{Acknowledgment}
\newacronym{adc}{ADC}{Analog to Digital Converter}
\newacronym{agv}{AGV}{Automated Guided Vehicle}
\newacronym{ai}{AI}{Artificial Intelligence}
\newacronym{aimd}{AIMD}{Additive Increase Multiplicative Decrease}
\newacronym{am}{AM}{Acknowledge Mode}
\newacronym{amc}{AMC}{Adaptive Modulation and Coding}
\newacronym{aqm}{AQM}{Active Queue Management}
\newacronym{awgn}{AWGN}{Additive White Gaussian Noise}
\newacronym{balia}{BALIA}{Balanced Link Adaptation}
\newacronym{bdp}{BDP}{Bandwidth-Delay Product}
\newacronym{bf}{BF}{Beamforming}
\newacronym{bo}{BO}{Back-Off}
\newacronym{bs}{BS}{Base Station}
\newacronym{bsr}{BSR}{Buffer Status Report}
\newacronym{cb}{CB}{Contention-Based}
\newacronym{cc}{CC}{Congestion Control}
\newacronym{cdf}{CDF}{Cumulative Distribution Function}
\newacronym{ce}{CE}{Control Element}
\newacronym{c-its}{C-ITS}{Connected Intelligent Transportation System}
\newacronym{cn}{CN}{Core Network}
\newacronym{cp}{CP}{Control Plane}
\newacronym{crs}{CRS}{Cell Reference Signal}
\newacronym{cs}{CS}{Congestion Control}
\newacronym{csi}{CSI}{Channel State Information}
\newacronym{csirs}{CSI-RS}{Channel State Information - Reference Signal}
\newacronym{d2d}{D2D}{Device-to-Device}
\newacronym{das}{DAS}{Distributed Antenna System}
\newacronym{dc}{DC}{Dual Connectivity}
\newacronym{dci}{DCI}{Downlink Control Information}
\newacronym{dl}{DL}{Downlink}
\newacronym{dmr}{DMR}{Deadline Miss Ratio}
\newacronym{dmrs}{DMRS}{DeModulation Reference Signal}
\newacronym{ds}{DS}{Dynamic Scheduling}
\newacronym{dft}{DFT}{Discrete Fourier Transform}
\newacronym{e2e}{E2E}{End-to-End}
\newacronym{e2el}{E2EL}{End-to-End Latency}
\newacronym{edca}{EDCA}{Enhanced Distribution Channel Access}
\newacronym{ecn}{ECN}{Explicit Congestion Notification}
\newacronym{edf}{EDF}{Earliest Deadline First}
\newacronym{embb}{eMBB}{Enhanced Mobile BroadBand}
\newacronym{endc}{EN-DC}{E-UTRAN-NR Dual Connectivity}
\newacronym{epc}{EPC}{Evolved Packet Core}
\newacronym{es}{ES}{Edge Server}
\newacronym{fdd}{FDD}{Frequency Division Duplexing}
\newacronym{fdma}{FDMA}{Frequency Division Multiple Access}
\newacronym{fl}{FL}{Federated Learning}
\newacronym{fs}{FS}{Fast Switching}
\newacronym{fr1}{FR1}{Frequency Range 1}
\newacronym{fr2}{FR2}{Frequency Range 2}
\newacronym{ftp}{FTP}{File Transfer Protocol}
\newacronym{gnb}{gNB}{Next Generation Node Base}
\newacronym{harq}{HARQ}{Hybrid Automatic Repeat reQuest}
\newacronym{hh}{HH}{Hard Handover}
\newacronym{hetnet}{HetNet}{Heterogeneous Network}
\newacronym{hol}{HOL}{Head-of-Line}
\newacronym{hqf}{HQF}{Highest-quality-first}
\newacronym{iab}{IAB}{Integrated Access and Backhaul}
\newacronym{ia}{IA}{Initial Access}
\newacronym{ibi}{IBI}{Instantaneous Buffer Information}
\newacronym{ieee}{IEEE}{Institute of Electrical and Electronics Engineers}
\newacronym{imt}{IMT}{International Mobile Telecommunication}
\newacronym{iot}{IoT}{Internet of Things}
\newacronym{iiot}{IIoT}{Industrial Internet of Things}
\newacronym{kpi}{KPI}{Key Performance Indicator}
\newacronym{ldpc}{LDPC}{Low-Density Parity Check}
\newacronym{lte}{LTE}{Long Term Evolution}
\newacronym{los}{LOS}{Line-of-Sight}
\newacronym{m2m}{M2M}{Machine to Machine}
\newacronym{mac}{MAC}{Medium Access Control}
\newacronym{mab}{MAB}{Multi-Armed Bandit}
\newacronym{mc}{MC}{Multi-Connectivity}
\newacronym{mcs}{MCS}{Modulation and Coding Scheme}
\newacronym{mec}{MEC}{Mobile Edge Cloud}
\newacronym{mi}{MI}{Mutual Information}
\newacronym{mimo}{MIMO}{Multiple Input Multiple Output}
\newacronym{mmwave}{mmWave}{Millimeter Wave}
\newacronym{mptcp}{MPTCP}{Multipath TCP}
\newacronym{mr}{MR}{Maximum Rate}
\newacronym{mrdc}{MR-DC}{Multi-Radio Access Technology Dual Connectivity}
\newacronym{mss}{MSS}{Maximum Segment Size}
\newacronym{mtu}{MTU}{Maximum Transmission Unit}
\newacronym{nfv}{NFV}{Network Function Virtualization}
\newacronym{nlos}{NLOS}{Non-Line-of-Sight}
\newacronym{noma}{NOMA}{Non-Orthogonal Multiple Access}
\newacronym{nsa}{NSA}{Non-Stand Alone}
\newacronym{nr}{NR}{New Radio}
\newacronym{ofdm}{OFDM}{Orthogonal Frequency Division Multiplexing}
\newacronym{ofdma}{OFDMA}{Orthogonal Frequency Division Multiple Access}
\newacronym{pbch}{PBCH}{Physical Broadcast Channel}
\newacronym{pdcch}{PDCCH}{Physical Downlink Control Channel}
\newacronym{pdcp}{PDCP}{Packet Data Convergence Protocol}
\newacronym{pdsch}{PDSCH}{Physical Downlink Shared Channel}
\newacronym{pdu}{PDU}{Protocol Data Unit}
\newacronym{phy}{PHY}{PHYsical}
\newacronym{plc}{PLC}{Programmable Logic Controller}
\newacronym{prr}{PRR}{Packet Reception Ratio}
\newacronym{prb}{PRB}{Physical Resource Block}
\newacronym{pss}{PSS}{Primary Synchronization Signal}
\newacronym{pucch}{PUCCH}{Physical Uplink Control Channel}
\newacronym{pusch}{PUSCH}{Physical Uplink Shared Channel}
\newacronym{qos}{QoS}{Quality of Service}
\newacronym{ra}{RA}{Random Agent}
\newacronym{ran}{RAN}{Radio Access Network}
\newacronym{rat}{RAT}{Radio Access Technology}
\newacronym{rb}{RB}{Resource Block}
\newacronym{red}{RED}{Random Early Detection}
\newacronym{rf}{RF}{Radio Frequency}
\newacronym{rl}{RL}{Reinforcement Learning}
\newacronym{rlc}{RLC}{Radio Link Control}
\newacronym{rlf}{RLF}{Radio Link Failure}
\newacronym{ris}{RIS}{Reconfigurable Intelligent Surface}
\newacronym{rr}{RR}{Round Robin}
\newacronym{rrc}{RRC}{Radio Resource Control}
\newacronym{rrm}{RRM}{Radio Resource Management}
\newacronym{rss}{RSS}{Received Signal Strength}
\newacronym{rsrp}{RSRP}{Reference Signal Received Power}
\newacronym{rsrq}{RSRQ}{Reference Signal Received Quality}
\newacronym{rtt}{RTT}{Round Trip Time}
\newacronym{rnti}{RNTI}{Radio Network Temporary Identifier}
\newacronym{sb}{SB}{Switch Buffer}
\newacronym{sc}{SC}{Single Carrier}
\newacronym{sch}{SCH}{Secondary Cell Handover}
\newacronym{se}{RE}{Resource Efficiency}
\newacronym{si}{SI}{Study Item}
\newacronym{sib}{SIB}{Secondary Information Block}
\newacronym{srs}{SRS}{Sounding Reference Signal}
\newacronym{sss}{SSS}{Secondary Synchronization Signal}
\newacronym{sinr}{SINR}{Signal to Interference plus Noise Ratio}
\newacronym{sla}{SLA}{Service Level Agreement}
\newacronym{sps}{SPS}{Semi-Persistent Scheduling}
\newacronym{spsch}{PSSCH}{Physical Sidelink Shared Channel}
\newacronym{spsf}{PSF}{Physical Sidelink Feedback}
\newacronym{sdu}{SDU}{Service Data Unit}
\newacronym{snpn}{SNPN}{Standalone Non-Public Network}
\newacronym{snr}{SNR}{Signal-to-Noise Ratio}
\newacronym{su}{SU}{Scheduling Unit}
\newacronym{tcp}{TCP}{Transmission Control Protocol}
\newacronym{ts}{TS}{Thompson Sampling}
\newacronym{tsa}{TS-A}{Thompson Sampling Agent}
\newacronym{tdd}{TDD}{Time Division Duplexing}
\newacronym{tdma}{TDMA}{Time Division Multiple Access}
\newacronym{tti}{TTI}{Transmission Time Interval}
\newacronym{ttt}{TTT}{Time-to-Trigger}
\newacronym{udp}{UDP}{User Datagram Protocol}
\newacronym{ue}{UE}{User Equipment}
\newacronym{up}{UP}{User Plane}
\newacronym{upa}{UPA}{Uniform Planar Array}
\newacronym{urllc}{URLLC}{Ultra-Reliable Low-Latency Communication}
\newacronym{us}{US}{Uplink Scheduler}
\newacronym{v2i}{V2I}{Vehicle-to-Infrastructure}
\newacronym{v2n}{V2N}{Vehicle-to-Network}
\newacronym{v2v}{V2V}{Vehicle-to-Vehicle}
\newacronym{v2x}{V2X}{Vehicle-To-Everything}
\newacronym{vm}{VM}{Virtual Machine}
\newacronym{wbf}{WBF}{Wired Bias Function}
\newacronym{wf}{WF}{Wired-first}
\newacronym{wlan}{WLAN}{Wireless Local Area Network}
\newacronym{wpt}{WPT}{Wireless Power Transfer}
\newacronym{ws}{WS}{Wired Scheduler}
\newacronym{wsa}{WS-A}{Wired Scheduler Agent}
\newacronym{xr}{XR}{Extended Reality}
\newacronym{mdp}{MDP}{Markov Decision Process}
\newacronym{tl}{TL}{Table-Less}
\newacronym{tb}{TB}{Table-Based}

\begin{document}
\title{Performance Analysis of Multi-Hop Networks \\ at Terahertz Frequencies}

\author{
    \IEEEauthorblockN{ 
        Sara Cavallero\IEEEauthorrefmark{1},
        Andrea Pumilia\IEEEauthorrefmark{1}, Giampaolo Cuozzo\IEEEauthorrefmark{1},
        Alessia Tarozzi\IEEEauthorrefmark{1}\IEEEauthorrefmark{2}, Chiara Buratti\IEEEauthorrefmark{1}\IEEEauthorrefmark{2} and Roberto Verdone\IEEEauthorrefmark{1}\IEEEauthorrefmark{2}
    }
    \IEEEauthorblockA{\IEEEauthorrefmark{1} National Laboratory of Wireless Communications (WiLab), CNIT, Italy}
    \IEEEauthorblockA{\IEEEauthorrefmark{2} Department of Electrical, Electronics, and Information Engineering "Guglielmo Marconi", University of Bologna, Italy}
}

\maketitle

\begin{abstract}
The emergence of THz (Terahertz) frequency wireless networks holds great potential for advancing various high-demand services, including Industrial Internet of Things (IIoT) applications. These use cases benefit significantly from the ultra-high data rates, low latency, and high spatial resolution offered by THz frequencies. However, a primary well-known challenge of THz networks is their limited coverage range due to high path loss and vulnerability to obstructions. 
This paper addresses this limitation by proposing two novel multi-hop protocols, Table-Less (TL) and Table-Based (TB), respectively, both avoiding centralized control and/or control plane transmissions. Indeed, both solutions are  distributed, simple, and rapidly adaptable to network changes. Simulation results demonstrate the effectiveness of our approaches
, as well as revealing interesting trade-offs between TL and TB routing protocols, both in a real IIoT THz network and under static and dynamic conditions.
\end{abstract}

\begin{IEEEkeywords}
Terahertz (THz), Industrial Internet of Things (IIoT), Medium Access Control (MAC), Table-Less Multi-Hopping, Table-Based Multi-Hopping.
\end{IEEEkeywords}

\IEEEpeerreviewmaketitle

\section{Introduction} 
\label{intro} 

The advent of mobile radio networks operating in the THz (Terahertz) band, which spans frequencies from 0.1 to 10 THz, promises to unleash the development of innovative applications \cite{akyildiz20206g}. The vast bandwidth available in these networks can offer ultra-high data rates, low latency, device miniaturization, and high spatial resolution \cite{rappaport2019wireless, chen2021terahertz}. These capabilities are particularly beneficial in challenging scenarios such as \gls{xr}, digital twins, and \gls{iiot} applications \cite{jornet2024evolution}. For instance, establishing wireless interconnections between virtualized \glspl{plc} and sensors/actuators on industrial machines necessitates per-user data rates of up to 100 Mbps \cite{etsigr001}, resulting in network throughputs of tens of Gbps even for a relatively small number of devices, along with latencies below 0.1 ms, highlighting the need for advancements in current wireless technologies.


However, one of the primary limitations of THz wireless networks is their restricted coverage range due to high path loss and susceptibility to obstacles \cite{moldovan2017coverage, shafie2020coverage}. A viable option to overcome this limitation is the use of multi-hop communication to extend the coverage area and ensure reliable connectivity. This approach is particularly important in environments with numerous obstructions and distances spanning several tens of meters, such as industrial plants. In particular, multi-hop networks leverage intermediate devices to relay data from the source to the destination, thereby overcoming the distance constraints inherent to THz frequencies.

The concept of multi-hop communication has been extensively studied in the context of wireless mesh networks \cite{alotaibi2012survey,WMN}. Several approaches have been proposed to address the challenges of routing in THz networks, including methods to optimize path selection, relay distribution, enhance reliability, and reduce latency \cite{xia2021multi, lou2023coverage, cavallero2023multi}. However, traditional routing protocols often involve complex algorithms and require significant overhead for maintaining routing tables and network state information \cite{xia2019link}. In addition, there exist many works dealing with energy-efficient routing metric, aiming at minimizing the packet forwarding energy consumption, without accounting for the lifetime of routers, that is assuming they are always on and ready for forwarding data (see, e,g., \cite{Energy_MANET}).

Given the dynamic evolution of the THz channel, which is highly susceptible to environmental changes due to the millimeter-scale wavelength, current approaches may encounter practical issues in real-world scenarios to quickly self-adapt the routing decisions.

This paper proposes two multi-hop approaches for THz mobile radio networks, that distinguish themselves from existing solutions through its simplicity and fast adaptability to dynamic conditions. Unlike conventional multi-hop schemes that rely on centralized control and/or control plane transmissions, we consider two fully distributed approaches that depend solely on ongoing user plane data transmissions to enable multi-hopping. In addition, in contrast to existing works, we account for the possibility that routers cannot forward data generated by neighbors, because of not being in reception state during data transmission.

Specifically, we propose \gls{tl} and \gls{tb} multi-hop network protocols: in the former, devices leverage reception phases for data collection and forwarding, while in the latter, users exploit neighbor tables to select the next hop based on the ongoing data transmission flow.

The proposed analysis considers an Unslotted-Aloha protocol at the \gls{mac} layer, motivated by the following main characteristics of THz frequencies: (i) the small coverage range of each device, which spatially limits collisions, and (ii) the long propagation delays compared to transmission times, which vary from link to link and reduce the probability of simultaneous reception of data at the receiver side. 


After demonstrating the need for multi-hop solutions over single-hop Unslotted Aloha, the simulation campaign analyzes the trade-offs between \gls{tl} and \gls{tb} routing, evaluating their performance in a real \gls{iiot} environment under both static and dynamic conditions, and demonstrating that the stringent requirements of \gls{iiot} applications can be achieved.

The paper is thus structured as follows. Sec. \ref{sec:system_model} introduces the system model, covering the scenario, traffic, and channel model. Sec. \ref{sec:mac_layer} details the \gls{mac} layer, with a brief recap of the basic Unslotted Aloha working principle. Then, the proposed \gls{tl} and \gls{tb} multi-hop approaches are explained in Sec. \ref{sec:multi_hop_approach}. The metrics used to assess our proposal are listed in Sec. \ref{sec:kpis}, and Sec. \ref{sec:numerical_results} presents the simulation results. Finally, in Sec. \ref{sec:conclusion}, we summarize the main findings of our paper.

\section{System Model}
\label{sec:system_model}


\subsection{Scenario}
\begin{figure}[!t]
  \begin{center}
    \includegraphics[width=0.9\columnwidth]{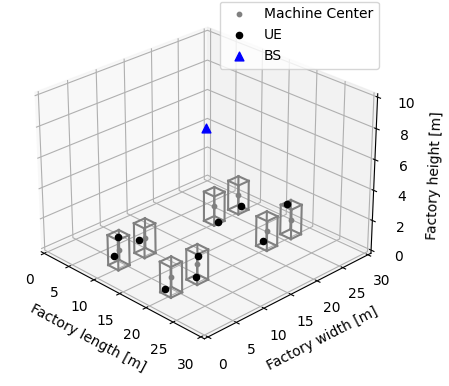}
    \caption{The considered \gls{iiot} scenario.}
    \label{fig:scenario}
  \end{center}
\end{figure}

Among the possible applications of THz frequencies, we focus on an \gls{iiot} scenario involving the remotization of \glspl{plc}, which represents one of the most challenging use cases \cite{etsigr001}. In particular, the simulated environment emulates an actual Italian plant measuring 52.36 x 35.4 x 8.5 meters. Within this plant, $O = 8$ packing machines of various sizes are positioned according to their actual locations. For simplicity, these industrial machines are modeled as cubes with sides of either 2 meters (i.e., small machines) or 4 meters (i.e., large machines).

We then consider $N$ \glspl{ue} (i.e., microcontroller boards embedding sensors) that are randomly distributed within the $O$ machines. These \glspl{ue} need to communicate in real-time with a remote \gls{plc}. To reach the latter, a \gls{bs} is located at the center of the top base of the factory, as shown in Fig. \ref{fig:scenario}. 


\subsection{Traffic Model}

This paper aims to analyze the achievable network performance in a worst-case scenario in terms of offered traffic. This scenario arises when the buffers of the \glspl{ue} are consistently full, meaning the \glspl{ue} always have a new DATA packet ready for transmission to the \gls{bs} immediately after receiving the \gls{ack} for the previous transmission. This analysis is crucial for evaluating the maximum network throughput of a THz network (with queue always full), as this metric depends on the type of traffic. Therefore, in the following sections, we will assume that all \glspl{ue} generate traffic with a fixed DATA packet size $P$.

\subsection{Channel Model}
\begin{figure}[!t]
  \begin{center}
    \includegraphics[width=\columnwidth]{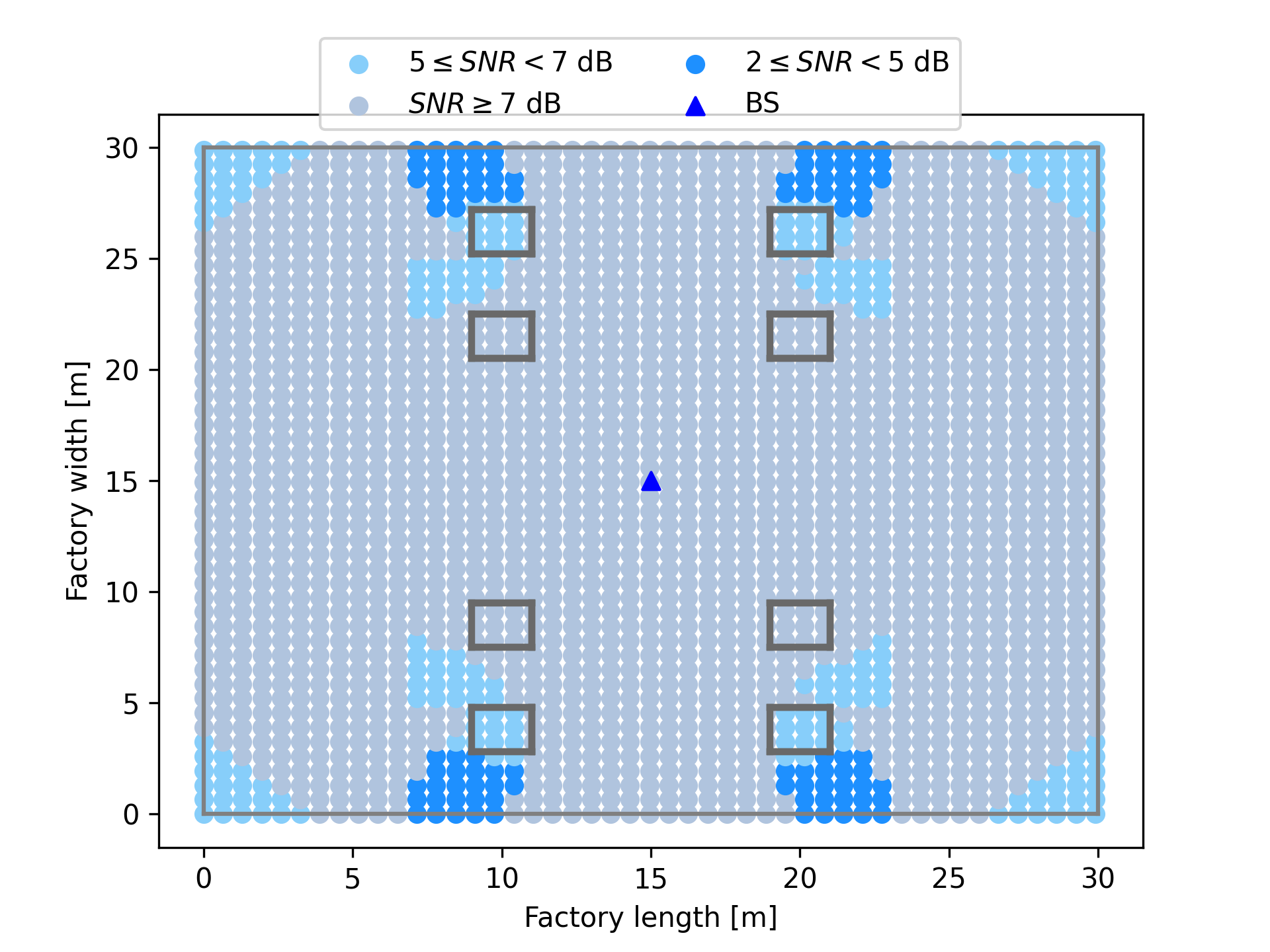}
    \caption{2D map illustrating the uplink $SNR$ distribution in the considered industrial plant.}
    \label{fig:coverage_map}
  \end{center}
\end{figure}

The channel propagation conditions considered in this work follow the Indoor Factory (InF) scenario of 3GPP TR 38.901
\cite{3gpp38901}. In particular, the channel is modeled with a narrowband description where the path loss $PL$ in dB can be formulated as follows:  
 \begin{equation}
     PL [dB] = \beta + \alpha \log_{10}(d [m]) + \gamma \log_{10}(f_c [GHz]), 
 \end{equation}
where $d$ is the Tx-Rx Euclidean distance, $f_c$ represents the carrier frequency, whereas $\alpha$, $\beta$, and $\gamma$ are real-values that depend on:

\begin{enumerate}
    \item \gls{los}/\gls{nlos} conditions that are computed deterministically based on the reciprocal positions of \glspl{ue} and \gls{bs}, where an intra-machine link is assumed to be in \gls{nlos};
    \item the clutter density of the scenario, either sparse (S) or dense (D). Specifically, the clutter density is defined as the ratio between the sum of the area occupied by the $O$ machines, and the area of the industrial plant. For the considered \gls{iiot} scenario, this results in a sparse clutter density;
    \item the height of the transmitter and receiver, either low (L), if both of them are below the height of the machines, or high (H), otherwise.
\end{enumerate}

The Signal-to-Noise Ratio ($SNR$) can be then expressed as:
\begin{equation}
 \begin{split}
     SNR \; [dB] &= P_{\rm TX} \; [dBW] + \eta_{\rm TX} \; [dB] + \eta_{\rm RX} \; [dB] + \\ &G_{\rm TX} \; [dB] + G_{\rm RX} \; [dB] - PL \; [dB] - P_{\rm N} \; [dBW],
     \label{snr_db}
 \end{split}
\end{equation}
where $P_{\rm TX}$ is the transmitted power, $\eta_{\rm TX}$, $\eta_{\rm RX}$ are the transmitter and receiver antenna efficiencies, respectively, $G_{\rm TX}$, $G_{\rm RX}$ are the transmitter and receiver gains, $PL$ is the path loss, and $P_{\rm N}=10\,\mathrm{log}_{\rm 10}(kT_0F_{\rm RX}B)$ is the noise power assuming the antenna temperature equal to the reference temperature $T_0$, with $k$ being the Boltzmann constant, $F_{\rm RX}$ is the receiver noise figure, and $B$ the bandwidth.
In particular, successful decoding occurs when $SNR$ overcomes a given threshold $SNR_{\rm TH}$, that is, $SNR \geq SNR_{\rm TH}$.

To provide a quantitative analysis, Fig. \ref{fig:coverage_map} presents the 2D coverage map of the industrial plant, illustrating uplink $SNR$ values based on the parameters in Table \ref{tab:simulation_params}. A typical threshold of $SNR_{\rm TH} \geq 7$ dB does not ensure full coverage, particularly under \gls{nlos} conditions, such as when \glspl{ue} are deployed within $O$ machines. This underscores the necessity of multi-hop transmissions for THz communications in \gls{iiot} scenarios, which is the focus of this paper.

\section{MAC Layer}
\label{sec:mac_layer}

\subsection{MAC protocol}
\label{sec:unslotted_aloha_protocol}
\begin{figure}[!t]
    \begin{center}
    \includegraphics[width=\columnwidth]{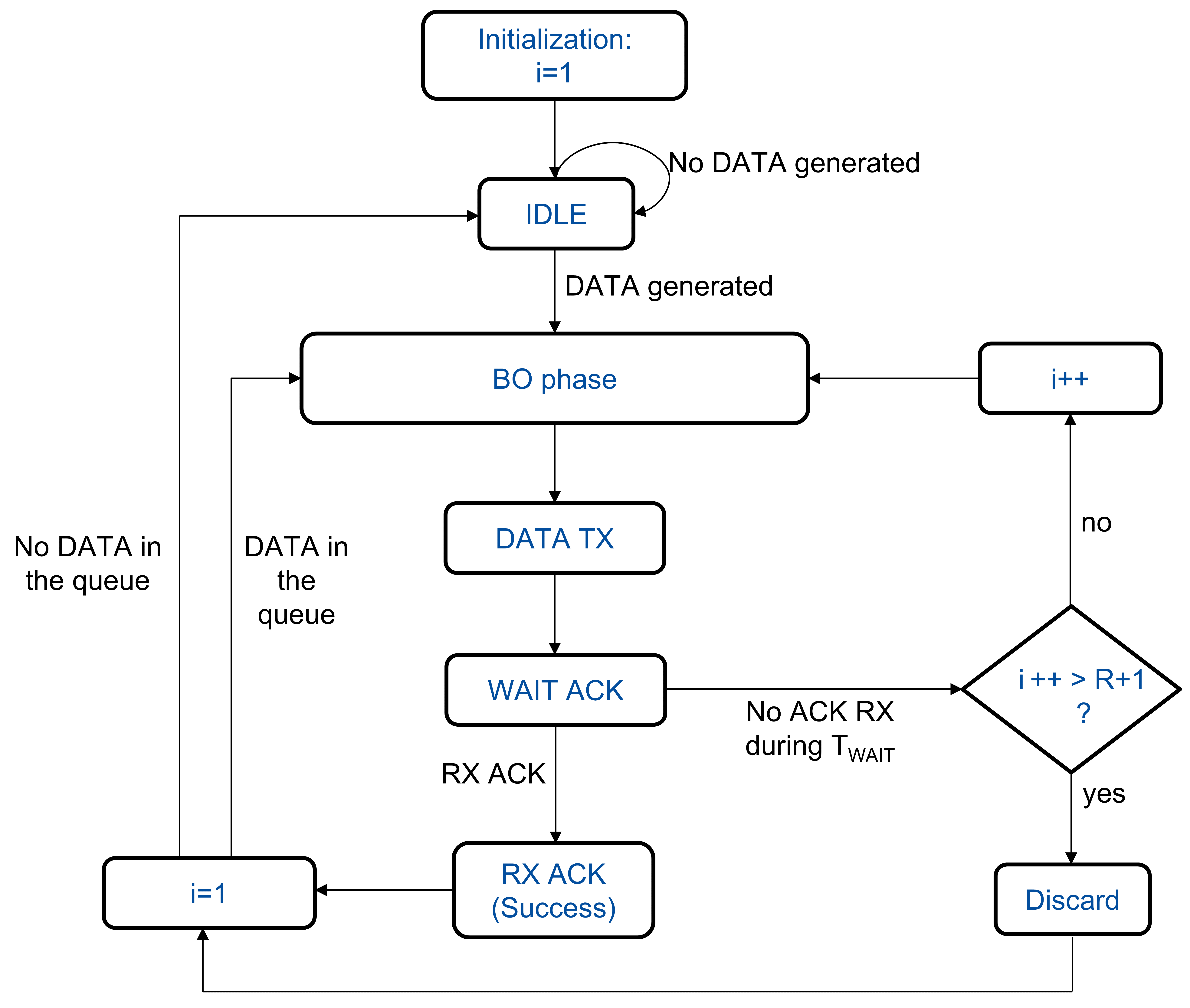}
    \caption{Flowchart of the \gls{mac} layer protocol from the \gls{ue} side.}
    \label{fig:unslotted_aloha_star_state_diagram}
    \end{center}
\end{figure}

As a trade-off between simplicity and performance \cite{ghafoor2020mac}, and given the unique characteristics of THz communications that help mitigate collisions, we consider a simple \gls{mac} protocol based on Unslotted Aloha \cite{hercog2020communication}, incorporating a random \gls{bo} even in the first transmission to further reduce collisions.

In particular, the \gls{mac} protocol works as follows. Each \gls{ue} is initially in IDLE mode. When new DATA has to be transmitted, the \gls{ue} initiates a \gls{bo} for $T_{\rm B} = T_{\rm BO}\, \xi$, where $T_{\rm BO}$ is the minimum \gls{bo} duration; $\xi$ is a uniform random number in the range $[1; 2^i C]$, with $C$ being an integer defining the maximum duration of the contention window; and $i$ being an integer counting the number of transmission attempts (starting at 1 for the first attempt). At the end of the \gls{bo} period, the \gls{ue} sends the DATA and enters reception mode for a maximum period, $T_{\rm WAIT}$. Specifically, we set $T_{\rm WAIT}=T_{\rm ACK}+2 \, \tau_{\rm p_{max}}$, since it is the time needed to transmit the DATA and receive the corresponding \gls{ack}, where $T_{\rm ACK}=\frac{8 \; P_{\rm A}}{R_{\rm b}}$, $P_{\rm A}$ is the number of bytes forming an \gls{ack}, $R_{\rm b} = B \log_2(M)$ is the bit rate, $M$ is the modulation order, and $\tau_{\rm p_{max}}$ is the maximum possible propagation delay in the considered \gls{iiot} scenario. 

If an \gls{ack} is received during $T_{\rm WAIT}$, the \gls{ue} either goes back to \gls{bo}, if a new DATA packet is in the \gls{mac} layer queue, or returns to IDLE if no DATA is queued. Conversely, if no \gls{ack} is received during $T_{\rm WAIT}$, the \gls{ue} retries the transmission up to a maximum number of attempts, $R$. The flowchart of the \gls{mac} layer protocol from the \gls{ue} side is shown in Fig. \ref{fig:unslotted_aloha_star_state_diagram}.

\subsection{Interference modeling}
Collisions are computed at the receiver side, taking into account the actual propagation delays from the transmitter(s). If two or more transmissions are received partially or completely overlapping in time, they are all considered lost, that is, we assume the worst-case scenario with no capture effect. Additionally, we assume a half-duplex mode of operation, meaning that if a device is busy transmitting an \gls{ack} and receives DATA simultaneously, the reception is discarded.

Note that we account for propagation delays, which vary for each \gls{bs}-\gls{ue} and \gls{ue}-\gls{ue} link, as they can be longer than transmission times at THz frequencies and thus impact the overall performance. Indeed, this peculiarity may actually reduce collision probability, since simultaneous transmissions (either DATA or \gls{ack}) may not lead to collisions if they are received at different times by the receiver.

\section{Multi-hop Approach}
\label{sec:multi_hop_approach}


In this section, we describe our proposal, that is, a \gls{tl} and \gls{tb} multi-hop algorithm at the network layer.


The key innovation of both approaches is that \glspl{ue} switch to reception mode during the \gls{bo} period, ensuring they can potentially receive DATA. Consequently, after \gls{bo}, \glspl{ue} transmit both their own DATA and any DATA received from other \glspl{ue}.

To realize the above approach, we set the \gls{bo} duration equal to $T_{\rm B} = T_{\rm DATA} + \tau_{\rm p_{\rm max}} + T_{\rm BO}\, \xi$, where $T_{\rm DATA}  = \frac{8 \; P}{R_{\rm b}}$ is the time needed to transmit a DATA, and $P$ is the number of bytes forming the DATA. If during \gls{bo} the \glspl{ue} correctly receive DATA, and the number of DATA in the \gls{mac} layer queue is below its capacity $Q$, they will enqueue it. In this case, \glspl{ue} will immediately acknowledge successful reception with an \gls{ack}. Note that, during the WAIT phase, each \gls{ue} remains in reception mode to receive the \gls{ack} for all transmitted DATA. This also allows a \gls{ue} to receive DATA from other \glspl{ue} during $T_{\rm WAIT}$ and forward them accordingly. 

\subsection{Table-less multi-hopping}
\label{sec:tl_mualoha}
In the \gls{tl} solution, after the \gls{bo} period, each \gls{ue} always sends its own DATA with a broadcast address, thereby producing the well-known broadcast storm problem \cite{alotaibi2012survey} but, at the same time, increasing the likelihood that at least one \gls{ue} within its transmission range is listening to the channel (i.e., is in reception mode during the \gls{bo} or WAIT phase) and can subsequently forward the DATA to the \gls{bs}. Then, \glspl{ue} enter the reception mode either (i) for a maximum duration of $T_{\rm WAIT}$ or (ii) until they correctly receive the \gls{ack} before $T_{\rm WAIT}$ expires. It is important to note that the non-negligible propagation delays, which contribute to reducing collisions and packet loss, allow us to account for \gls{ack} transmissions from \glspl{ue} that have successfully received packets for forwarding. This approach contrasts with conventional flooding methods in the literature, where such delays are often overlooked \cite{Flooding}.

\subsection{Table-based multi-hopping}
\label{sec:tb_mualoha}
In the \gls{tb} approach, \glspl{ue} create a neighbor table containing for each other \gls{ue}: (i) an indication of whether it has the \gls{bs} in its own table, (ii) the number of \glspl{ack} received from it; and (iii) the corresponding $SNR$s. Hence, after the \gls{bo} phase, each \gls{ue} transmits its own DATA with a unicast address to the optimal receiver selected based on the neighbor table. Specifically, the optimal receiver is always the \gls{bs} if it is present in the neighbor table; otherwise, it is the \gls{ue} that has the \gls{bs} in its table and/or from which the highest number of \glspl{ack} has been received at the highest $SNR$.

When the neighbor table is empty, after \gls{bo}, the \glspl{ue} initiate a discovery phase without adding any control plane overhead. Indeed, \glspl{ue} simply transmit its own DATA in broadcast and then enter reception mode for $T_{\rm WAIT}$. During this time, they populate the neighbor table based on the \glspl{ack} received from all receivers that have correctly collected the broadcast transmission
. Once the neighbor table contains at least one entry, the discovery phase concludes. This means that, in the next \gls{bo} phase, DATA are sent in unicast to the optimal receiver and the WAIT phase ends before $T_{\rm WAIT}$ if the correct \gls{ack} is received (as in the \gls{tl} case).

Note that each \gls{ue} removes an entry from the table after a Time To Live, $TTL$, expires, which is defined as the number of consecutive missed receptions of an \gls{ack} from a specific neighbor.

\subsection{Final remarks}
For both \gls{tb} and \gls{tl} approaches, it is important to note that:

\begin{enumerate}
    \item During $T_{\rm WAIT}$, \glspl{ue} can also receive DATA from other \glspl{ue}. In this case, they will transmit the corresponding \gls{ack}(s) at the end of the WAIT phase;
    \item The \gls{bs} discards any duplicate DATA generated by the same \gls{ue} but forwarded by different \glspl{ue};
    \item \glspl{ue} discard any received DATA that (i) is already present in their queue, (ii) originates from themselves (to prevent loops), or (iii) exceeds the predefined hop count limit $H$ (i.e., the maximum number of relays to be traversed before reaching the \gls{bs});
    \item The \glspl{ue} discard any \gls{ack} received during \gls{bo} or WAIT but not intended for them.
\end{enumerate}

\section{Key Performance Indicators}
\label{sec:kpis}
In this section, we present the \glspl{kpi} that have been utilized to validate the proposed multi-hop approaches (see Sec. \ref{sec:multi_hop_approach}) within our THz-based system model (see Sec. \ref{sec:system_model}).

\subsection{Success probability}
The success probability, $p_{\rm s}$, is defined as the ratio between the average number of DATA successfully received at the \gls{bs}, and the total number of DATA transmitted by the \glspl{ue}, that is,

\begin{equation}
\label{eq:success_probability}
    p_{\rm s} = \frac{1}{N}  \sum_{j=1}^N \frac{N_{\rm RX_j}}{N_{\rm TX_j}},
\end{equation}
where $N$ is the total number of \glspl{ue}, $N_{\rm RX_j}$ is the number of DATA successfully received at the \gls{bs} by the $j$-th \gls{ue}, and $N_{\rm TX_j}$ is the number of DATA transmitted by the $j$-th \gls{ue}. We then recall that a DATA packet is deemed successfully received at the \gls{bs} when its $SNR \geq SNR_{\rm TH}$ and it has not collided with any other transmission.



\subsection{Network Throughput}
\label{sec:network_throughput}
The network throughput, $S$, is defined as the number of information bits per second successfully received at the \gls{mac} layer of the \gls{bs}, that is,

\begin{equation}
    S = \frac{P N_{\rm R}}{T_{\rm S}},
\end{equation}
where $P$ is the DATA size, $N_{\rm R}$ is the number of DATA successfully received at the \gls{bs}, and $T_{\rm S}$ is the simulation time.

\subsection{Latency}
The average latency, $\overline{L}$, is defined as the average time needed by \glspl{ue} to transmit DATA with success, that is,

\begin{equation}
\label{eq:latency}
\overline{L} = \frac{1}{N} \sum_{j=1}^{N} \frac{1}{N_{\rm P_j}} \sum_{i=1}^{N_{\rm P_i}} L_{j,i}
\end{equation}
where $N_{\rm P_j}$ is the number of DATA generated by the $j$-th \gls{ue} and not discarded, and $L_{j,i}$ is the time interval from the generation of the $i$-th DATA by the $j$-th \gls{ue} to the reception of the corresponding \gls{ack}\footnote{Note that \glspl{ue} can receive \glspl{ack} from other \glspl{ue} acting as relays to the \gls{bs}.}. When the \gls{bs} successfully receives the DATA from the first relay, we calculate the latency for that packet, denoted as $L_{j,i}$, by summing 
$T_{\rm ACK}$ and the propagation delay of the \gls{ack}. It is important to note that the \gls{ack} from the \gls{bs} may not be received by the originating \gls{ue}, as it is not in reception mode after already receiving confirmation from the first relay.  

\section{Performance Evaluation}
\label{sec:numerical_results}

\begin{table}[!t]
	\renewcommand{\arraystretch}{1.1}
	\caption{Simulation parameters.}
	\label{tab:simulation_params}
	\centering
	\begin{tabular}{|c|c|c|}
		\hline
		\multicolumn{1}{|c|}{\textbf{Symbol}} & 
	  \multicolumn{1}{|c|}{\textbf{Description}}  & 
    	\multicolumn{1}{|c|}{\textbf{Value}}\\
     \hline
$f_{\rm c}$  & Carrier frequency & 100 GHz \\ \hline 
$B$ & Bandwidth & 25 GHz \\ \hline
$P_{\rm TX, UE}$ & Power transmitted from a \gls{ue} & 25 dBm \\ \hline
$P_{\rm TX, BS}$ & Power transmitted from the \gls{bs} & 30 dBm \\ \hline
$\eta_{\rm UE}$ & Antenna efficiency of a \gls{ue} & 0 dB \\ \hline
$\eta_{\rm BS}$ & Antenna efficiency of the \gls{bs} &  0 dB \\ \hline
$G_{\rm UE}$ & Antenna gain of a \gls{ue} & 8 dB \\ \hline
$G_{\rm BS}$ & Antenna gain of the \gls{bs} & 10 dB \\ \hline
$F_{\rm UE}$ & Noise figure of a \gls{ue} & 9 dB \\ \hline
$F_{\rm BS}$ & Noise figure of the \gls{bs} & 8 dB \\ \hline
$T_{\rm 0}$ & Reference temperature & 290 K \\ \hline
$M$ & Modulation order & 4 \\ \hline
$SNR_{\rm TH}$ & $SNR$ threshold & 7 dB \\ \hline
$C$  & Integer value defining the \gls{bo} period & 5 \\ \hline 
$R$  & Maximum number of retransmissions & 3 \\ \hline
$H$ & Maximum number of hops of a single DATA & 4 \\ \hline 
$P_{\rm A}$  & Size of an \gls{ack} & 10 B \\ \hline 
$P$  & Size of a DATA & 20 B \\ \hline 
$Q$ & Length of the \glspl{ue}' queue & \{5, 9\} \\ \hline
$TTL$ & Time To Live for an entry in the neighbor table & 3 \\ \hline
$T_{\rm BO}$  & \gls{bo} minimum time slot duration & 1.6 ns \\ \hline 
$N_{\rm S}$  & Number of simulations & 20 \\ \hline
$T_{\rm S}$ & Simulation time & 0.5 ms \\ \hline
	\end{tabular}
\label{parameter}
\end{table}

\subsection{Simulation setup}
Simulations parameters, if not otherwise specified, are reported in Table \ref{tab:simulation_params}. In particular, all results have been obtained by averaging over $N_{\rm S}$ simulations of duration $T_{\rm S}$, where each simulation mainly differs for the spatial distribution of \glspl{ue}. It is worth noting that \glspl{ue} are distributed randomly within the $O$ machines so that half of them are connected to the \gls{bs} (i.e., $SNR \geq SNR_{\rm TH}$), and the others are not. 

Based on the system model described in Sec. \ref{sec:system_model}, we first consider the single-hop Unslotted Aloha-based \gls{mac} layer protocol (UALOHA), as described in Sec. \ref{sec:unslotted_aloha_protocol} with no movements of the \glspl{ue}. This serves as a baseline to illustrate the limitations of single-hop communication and highlight the necessity of multi-hop approaches.

Beyond this first analysis, our simulation campaign has considered four cases:

\begin{enumerate}
    \item \textit{\gls{tl}, Static}: This is the proposed \gls{tl} multi-hop approach described in Sec. \ref{sec:tl_mualoha}, where \glspl{ue} do not change positions during a single simulation run;
    \item \textit{\gls{tl}, Dynamic}: This is similar to case 1 but with \glspl{ue} changing positions during a single simulation run. Specifically, the \glspl{ue} are moved to modify their potential neighbor(s);
    \item \textit{\gls{tb}, Static}: This is the proposed \gls{tb} multi-hop approach described in Sec. \ref{sec:tb_mualoha}, where \glspl{ue} occupy fixed positions during a single simulation run;
    \item \textit{\gls{tb}, Dynamic}: This is similar to case 3 but with \glspl{ue} changing positions during a single simulation run, as in case 2.
\end{enumerate}

\subsection{Numerical results}

\begin{table}[!t]
	\renewcommand{\arraystretch}{1.1}
	\caption{Success probability comparison between single-hop UALOHA, \gls{tb} and \gls{tl}, when considering static conditions.}
	\label{tab:ps_ualoha_TB_TL}
	\centering
	\begin{tabular}{|c|c|c|c|}
		\hline
		\multicolumn{1}{|c|}{\textbf{N}} & 
	  \multicolumn{1}{|c|}{\textbf{UALOHA}}  & 
    	\multicolumn{1}{|c|}{\textbf{\gls{tb}}} & \multicolumn{1}{|c|}{\textbf{TL}}\\
     \hline
$4$  & $0.5$ & $0.999$ & $0.997$ \\ \hline 
$6$  & $0.5$ & $0.999$ & $0.989$ \\ \hline
$8$  & $0.499$ & $0.998$ & $0.979$ \\ \hline
$10$  & $0.499$ & $0.991$ & $0.970$ \\ \hline
$12$  & $0.498$ & $0.988$ & $0.961$ \\ \hline
\end{tabular}
\label{parameter}
\end{table}

\begin{figure}[!t]
\begin{center}
\includegraphics[width=\columnwidth]{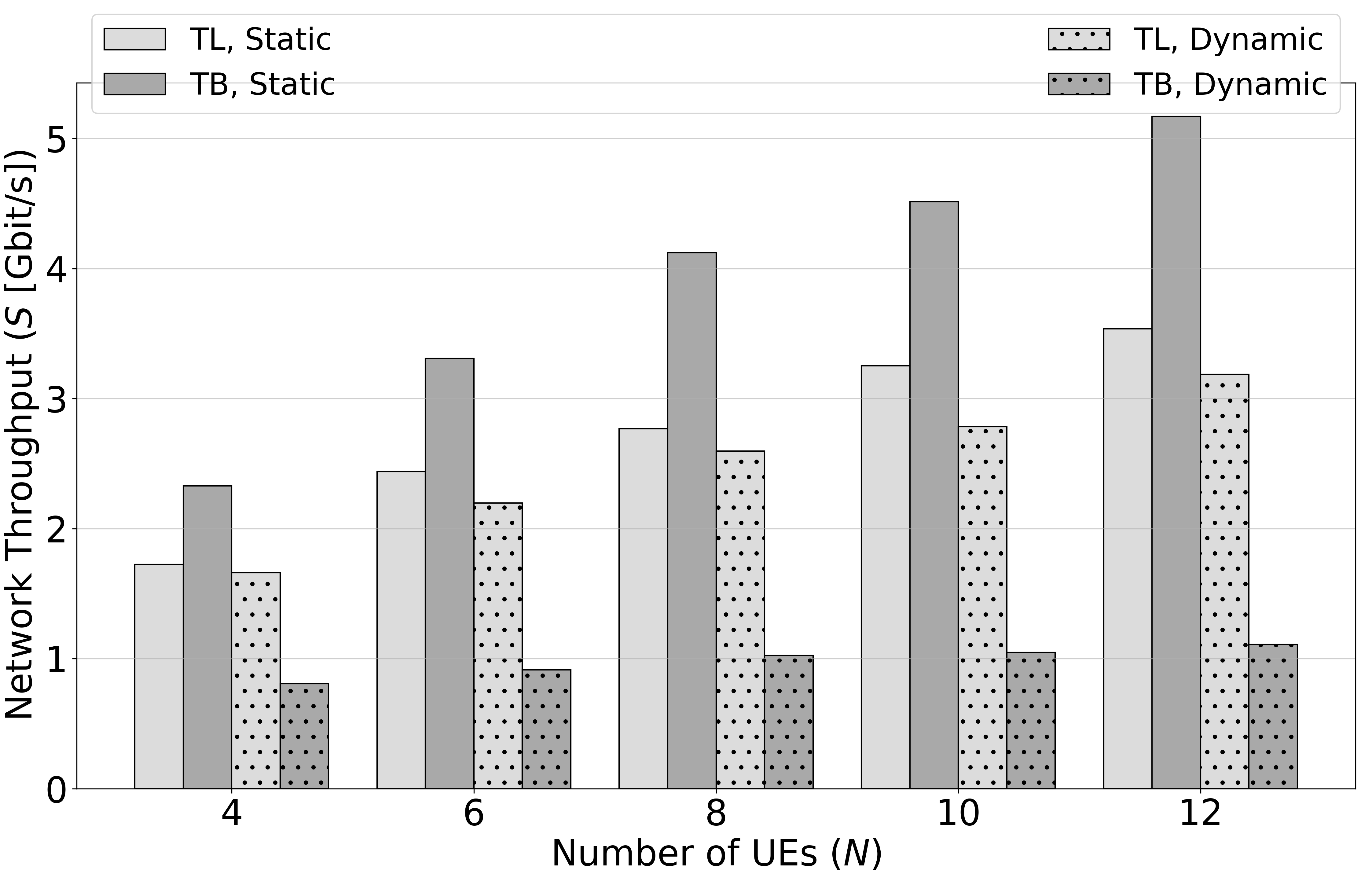}
\caption{Network Throughput, $S$, as a function of the number of \glspl{ue}, $N$, and the four simulation cases.}
\label{fig:throughput_vs_n}
\end{center}
\end{figure}

We start our analysis with Table \ref{tab:ps_ualoha_TB_TL}, which evaluates the success probability, $p_{\rm s}$, as a function of the number of \glspl{ue}, $N$, for the single-hop UALOHA protocol, and our proposals, namely \gls{tb}, and \gls{tl} routing protocols. In this case, we consider only static conditions. As expected, $p_{\rm s}$ decreases with increasing $N$, due to higher collision rates, and UALOHA exhibits the worst performance, justifying the need for multi-hopping. This is because half of the \glspl{ue} remain disconnected from the \gls{bs} due to spatial constraints in the considered \gls{iiot} scenario.  
In contrast, the proposed multi-hop extensions (\gls{tl} and \gls{tb}) enable full connectivity by allowing all \glspl{ue} to communicate with the \gls{bs}, even being outside its coverage range. This is achieved through broadcast or unicast transmissions characterizing the \gls{tl} and \gls{tb} versions, respectively. 
Additionally, \gls{tb} outperforms \gls{tl} due to its reduced overhead from unicast transmissions, which lowers the collision probability. 

After demonstrating the need for multi-hop communications, Figure \ref{fig:throughput_vs_n} presents the network throughput, $S$, as a function of the number of \glspl{ue}, $N$, across the four simulation cases. As observed, $S$ increases with $N$ since the total number of packets successfully received at the \gls{bs}, $N_{\rm R}$, also grows with $N$, despite a higher collision rate.  This results in a network throughput in the order of few Gbit/s, which satisfies the requirements of \gls{iiot} applications \cite{etsigr001}.
Furthermore, as previously shown, \gls{tb} outperforms \gls{tl} under static conditions due to a higher number of successfully received packets at the \gls{bs}. This improvement stems from the next-hop selection mechanism, which significantly reduces collisions and improves $p_{\rm s}$.
The reader can also note that \gls{tl} and \gls{tb} approaches behave differently under static and dynamic conditions. In fact, \gls{tl} is robust to changes in network topology due to \glspl{ue} broadcasting their DATA without selecting a specific next hop, thereby leveraging spatial redundancy at the expense of the known drawbacks caused by the broadcast storm problem \cite{alotaibi2012survey}. In contrast, \gls{tb} requires \glspl{ue} to update their neighbor tables for next-hop selection, which can lead to missed transmissions when the selected neighbor moves out of range, significantly lowering $N_{\rm R}$ and, consequently, network throughput $S$ in dynamic conditions. 

\begin{figure}[!t]
\begin{center}
\includegraphics[width=\columnwidth]{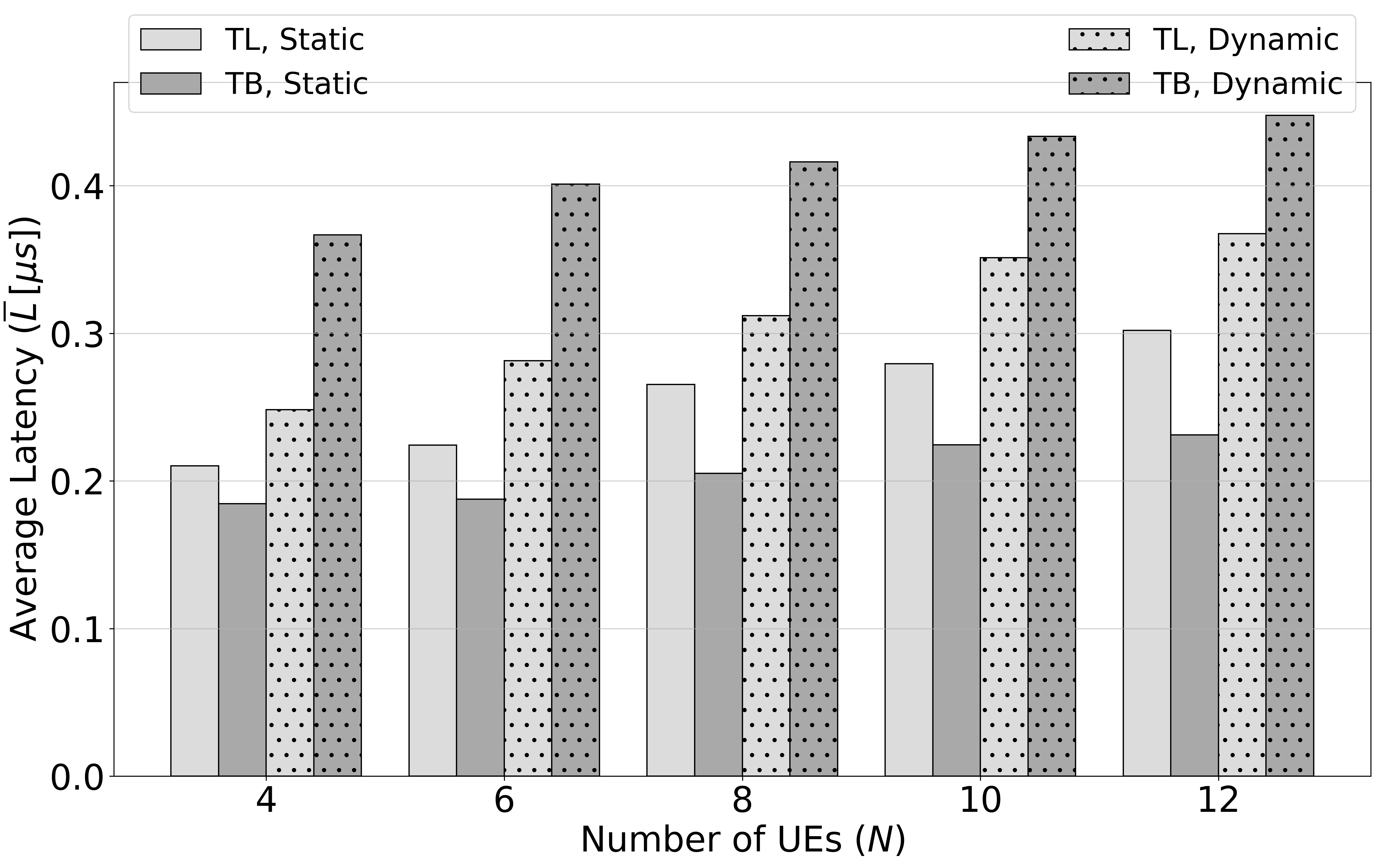}
\caption{Average latency, $\overline{L}$, as a function of the number of \glspl{ue}, $N$, and the four simulation cases.}
\label{fig:latency_vs_n}
\end{center}
\end{figure}

Then, Figure \ref{fig:latency_vs_n}, shows the Average Latency, $\overline{L}$, as a function of the number of \glspl{ue}, $N$, and again the four simulation cases. As expected, $\overline{L}$ increases with $N$ due to the higher number of collisions and retransmissions caused by the greater number of \glspl{ue} competing for the channel, while still remaining below 0.5 $\rm  \mu$s.  
These results further confirm the better performance of \gls{tb} under static conditions because of the higher reliability given by the choice of the next hop and the advantage of \gls{tl} in dynamic scenarios thanks to the broadcast transmission of packets.  

\begin{figure}[!t]
\begin{center}
\includegraphics[width=\columnwidth]{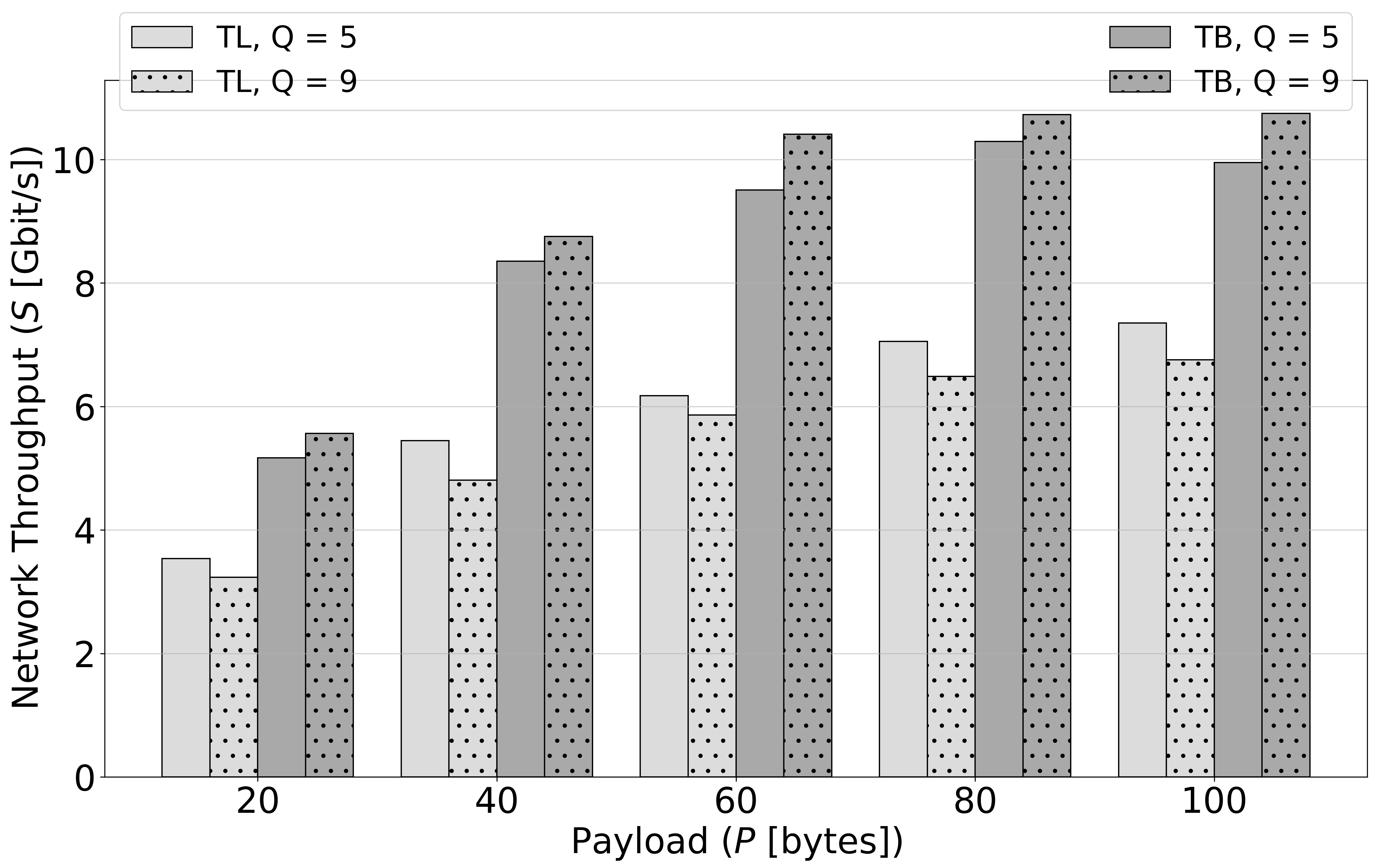}
\caption{Network Throughput $S$, as a function of DATA size $P$, the length of the \glspl{ue}' \gls{mac} layer queue $Q$, and \gls{tb} and \gls{tl} in the static case for a fixed number of $N = 12$ \glspl{ue}.}
\label{fig:S_vs_P}
\end{center}
\end{figure}

Finally, Figure \ref{fig:S_vs_P} illustrates the network throughput, $S$, as a function of the DATA size $P$, the \gls{mac} layer queue length $Q$, for a fixed number of $N = 12$ \glspl{ue}. The simulation cases analyzed are \gls{tb} and \gls{tl} under static conditions and it can be noted that, for both solutions, $S$ increases as $P$ grows from 20 to 100 bytes. Despite a lower success probability due to a higher number of collisions, these results confirm the robustness of the proposed solutions, demonstrating that throughput in the order of several Gbit/s can be achieved even with larger payloads. In particular, the traffic offered from each \gls{ue} increases with $P$ (ranging from 0.35 Gbit/s to 1.5 Gbit/s) and with that also the number of packets successfully received at \gls{bs}, $N_{\rm R}$. However, the increasing trend of $S$ tends to saturate as $P$ increases, as the interference effect starts to become dominant w.r.t the higher offered traffic.

It can also be observed that \gls{tl} and \gls{tb} have opposite behaviors when modifying the buffer length. 
This trend can be attributed to the robustness of \gls{tb} in the static scenario, where it maintains a high success rate, particularly with large queues. Although collisions occur, their impact remains limited, allowing a larger buffer to accomodate more successfully received packets at the \gls{bs}. In contrast, in TL, setting $Q=9$ leads to performance degradation due to excessive collisions, since all data are transmitted in broadcast. However, reducing $Q=5$ mitigates these collisions, resulting in improved performance.
Nevertheless, the best performance of \gls{tb} with respect to \gls{tl} in the static scenario is still confirmed.





\section{Conclusions}
\label{sec:conclusion}
In this paper, we proposed two \gls{tl} and \gls{tb} multi-hop approaches for THz wireless networks, based on the well-known Unslotted Aloha \gls{mac} layer protocol. We consider the unique properties of THz frequencies, such as small coverage range and link-by-link propagation delays, and we emulate a real \gls{iiot} environment under static and dynamic conditions. Through extensive simulations, we demonstrated the necessity of multi-hop approaches and showed that the proposed solutions significantly enhance network throughput -- reaching several Gbit/s -- while reducing the average latency to below 0.5 $\rm \mu$s. These promising results indicate the potential to meet the demanding requirements of \gls{iiot} applications.
Moreover, we show that under static conditions, the considered \gls{tb} network protocol provides better performance. In contrast, the lower overhead of the \gls{tl} algorithms is more advantageous in dynamic conditions, as maintaining routing tables can be costly. 
Future works will focus on leveraging artificial intelligence to optimize the identified trade-offs between \gls{tl} and \gls{tb} network algorithms.

\section*{Acknowledgments}
This work has been performed in the framework of the HORIZON-JU-SNS-2022 project TIMES, cofunded by the European Union. Views and opinions expressed are however those of the authors only and do not necessarily reflect those of the European Union.

\bibliographystyle{IEEEtran}
\bibliography{IEEEabrv,biblio}
\end{document}